\documentclass{aa}
\usepackage{graphicx}

\begin{document}

\title{The Star Cluster System of the NGC~7673 Starburst}

\author{Nicole Homeier \inst{1} \inst{2} \and John S. Gallagher III \inst{2} \and Anna Pasquali \inst{3}}


\institute{ESO, Karl Schwarzschild Str. 2, D-85748 Garching bei M\"{u}nchen \and 
University of Wisconsin, Department of Astronomy, 475 N. Charter St., Madison WI 53706, USA \and ESO ST-ECF, Karl Schwarzschild Str. 2, D-85748 Garching bei M\"{u}nchen }

\date{Received / Accepted}

\titlerunning{SCS of NGC 7673}
\authorrunning{Homeier, Gallagher, \& Pasquali}

\abstract{We investigate the star cluster system
in the starburst galaxy NGC~7673 using archival {\it Hubble Space Telescope}
 WFPC2 broad-band images. For the first time we are able
to examine the internal structures of the prominent optical clumps 
in this galaxy. The clumps are composed of young stars, $16-33 \%$ of which
are in bright star clusters. We identify 268 star cluster 
candidates in both the F555W and 
F814W images, and 50 clusters with the F255W
filter. These data allow us to estimate ages and masses using color-magnitude 
and two-color diagrams for our sample. We find a population
of young, $< 6$~Myr clusters located throughout
the galaxy with concentrations in the clumps. Star cluster mass estimates
are $5-50 \times 10^{4}$~M$_{\odot}$ for the brightest objects. The 
starburst remains active in physically well-separated 
regions, indicating a widespread starburst trigger. We discuss clump
lifetimes, their implications for the future evolution of NGC~7673, and
possible relationships to high redshift starbursts.}

\maketitle

\keywords{{Galaxies: Evolution, Galaxies: Starburst, Galaxies: 
Individual: NGC~7673, Galaxies: Interactions, Galaxies: Star Clusters}

\section{Introduction}

Starburst galaxies churn gas into stellar light at tremendous rates,
so fast that the time to exhaust this material is short compared to
the age of the universe (e.g., \cite{Mira92}).  In recent years they
have also become notorious for their propensity to form swarms of star
clusters, including massive, compact super star clusters (SSCs),
which have been suggested by many authors to be the progenitors of
globular clusters (e.g., \cite{HOG94}, \cite{WS95},
\cite{AZ01}).

The starburst phenomenon occurs rarely in the local universe, but
increases in frequency at larger lookback times
(e.g., \cite{H77}, \cite{Liu98}, \cite{Glaze99} and references
therein), which may be connected to the observed excess of distant
``faint blue galaxies'' (\cite{E97}).  Studies of moderate redshift
($z \sim 1$) galaxy samples reveal substantial populations of luminous
galaxies with high star formation rates (SFRs) (\cite{CHS95},
\cite{Oetal01}).  These include the blue ``compact narrow emission
line galaxies'' (CNELGs; \cite{Ketal94}, \cite{Ketal95},
Guzm\'{a}n et al. 1996, 1997, 1998, \cite{Petal97}).

Similar objects
have been categorized into various groups with various names, such as
the ``luminous blue compact galaxies'', or LBCGs (our preferred term
for such objects; \cite{Jetal01}). Galaxies of this class have small
linear sizes ($<$20~kpc), luminosities near or above $L^*$, high
surface brightnesses, strong emission lines, and blue colors, all
features that are indicative of enhanced SFRs.

The evolutionary fate of moderate redshift LBCGs is a subject of some
debate, and depends strongly on the current and future SFRs in
comparison with the extent of their gas reservoirs.  One
scenario holds that the most compact of the LBCGs will evolve into
spheroidal stellar systems covering a range in mass (e.g.,
\cite{BR92}, \cite{Getal98}).  The similarity of this subclass of
LBCGs to luminous, young, star-forming \ion{H}{ii} galaxies has led to
a classification of some systems as ``\ion{H}{ii}-like''
(\cite{Petal97}, \cite{Getal97}).  

Passive evolutionary models predict
that after 4-6 Gyr, the luminosities and surface brightnesses of these
\ion{H}{ii}-like LBCGs could resemble local low-mass ellipticals, like
NGC~205 (\cite{Getal98}). The viability of this scenario depends on
the star formation in LBCGs having a duration of less than 1~Gyr, and 
in that time
they must lose almost all their gas (\cite{Petal01}), which is
difficult to accomplish in any but the least massive dwarfs
(\cite{MLF99}, \cite{FT00}).

Alternatively, LBCGs could be analogs to intense starbursts in nearby
disk galaxies, especially when observed in near face-on orientations
that favor the escape of blue/ultraviolet light from the starburst
region (\cite{GHB89}, \cite{GCH00}, \cite{BvZ01}).  The
evolutionary scenario for disk-like LBCGs does not necessarily lead to
their descendants being spheroidal galaxies.  In nearby examples of
this starburst mode, relatively minor interactions between galaxies
can yield major starbursts in which the disk is only moderately
perturbed (e.g., M82).  If the post-burst systems have sufficient gas
supplies to allow star formation to continue, albeit at lower rates
than in the current bursts, then they can appear at the present epoch
as comparatively normal star-forming disk galaxies (\cite{Petal97},
\cite{Hetal00}).

Related to these evolutionary questions is the issue of bulge
formation.  Theoretical work has shown that gas-rich ($> 10 \%$ in
\ion{H}{i}) disks subject to gravitational instability may form large
clumps of gas, up to $10^{9}$ M$_{\odot}$, which rotate in the plane
of the disk (\cite{EKT93}, \cite{N99}). Such clumps experience strong star
formation, resulting in a morphologically peculiar galaxy, and could
appear like the 'clumpy' galaxy investigated here. 

The clumps suffer
dynamical friction from the surrounding visible and 
dark matter, leading them to spiral inwards and accumulate in the 
central region, potentially forming a bulge.  In this
scenario the disk forms {\it first}, hosts the initial round of star
formation, produces a bulge. The remaining 
gas could then fuel a more sedate course of evolution. However, a key factor 
in this model is whether the clumps can retain 
their identities over the $\sim$100~Myr time scales required for 
dynamical friction to act (\cite{N99}).

To clarify the general issues of how starbursts connect to star
formation processes and galaxy evolution, we concentrate on
understanding nearby starbursts whose internal properties are
observable.  With this goal in mind, the Wide Field Planetary Camera 2
(WFPC2) Investigation Definition Team GTO program included exploration
of small scale structures in blue starburst galaxies with M$_{B} < -18$
 and high optical surface brightness, many of which are members
of the LBCG class (\cite{GHC00}).  NGC~7673 (a.k.a. IV Zw 149, Markarian 325, UGC
12607, and UCM 2325+2318), a 'disturbed spiral' or 'clumpy irregular'
galaxy was chosen for a multi-wavelength imaging investigation
extending from the mid-ultraviolet to the near infrared. NGC~7673 is a
member of the LBCG class, and a luminous infrared source (e.g.,
\cite{GHB89}, \cite{Hetal89}, \cite{SM96}).  As such, it presents a
unique opportunity to understand the inner workings of a high SFR
object and its potential evolutionary connections to distant and
present-day galaxy populations.

This paper presents an analysis of star clusters in the clumpy
starburst regions of NGC~7673 based on archival images obtained with
WFPC2 on the {\it Hubble Space Telescope}.
We assume H$_{0} = 70$~km~s$^{-1}$~Mpc$^{-1}$; the
recession velocity of 3408~km~s$^{-1}$ from the NASA/IPAC
Extragalactic Database\footnote {The NASA/IPAC Extragalactic Database
(NED) is operated by the Jet Propulsion Laboratory, California
Institute of Technology, under contract with the National Aeronautics
and Space Administration.} for NGC~7673 then implies a distance of
49~Mpc, and a projected scale of 250 pc per arcsec. The distance
modulus for this galaxy is 33.45 and one WFC pixel covers a projected
scale of approximately 25 pc.  

The next section summarizes properties of this unusual galaxy.  Our
observations and photometry of star clusters are detailed in \S 3, and
\S 4 provides information on our choice of models for the spectral
evolution of the clusters.  In \S 5 we discuss the morphology of
NGC~7673 as seen with WFPC2, \S 6 discusses the star cluster colors 
and magnitudes in terms of their ages, 
and \S 7 briefly characterizes the star-forming clumps.
Our results are discussed \S 8,
and in \S 9 we present our conclusions.

\section{Properties of NGC~7673}

The overall structure and kinematics of NGC~7673 were discussed by
Homeier \& Gallagher (1999, hereafter HG), who argued that, despite the 
disturbed optical appearance of this system, it is a 
relatively unperturbed, rotating disk galaxy seen nearly face-on. 

Initial investigations of NGC~7673 highlighted its disturbed and
``clumpy'' optical appearance (\cite{ML71}, \cite{BK75}, \cite{CH76}),
and earned it a classification as a ``clumpy, irregular'' galaxy.
The spiral pattern that is prominent in our images also was noted
in earlier studies (\cite{dV76}, \cite{H77}, \cite{CHH82}). Optical
spectra revealed the presence of a broad emission component underlying
the narrow lines more commonly associated with \ion{H}{ii} regions,
and a remarkably constant radial velocity across the galaxy
(\cite{DA82}, \cite{TT87},
\cite{HG99}). 

\ion{H}{i} observations of NGC~7673 and its companion NGC~7677 ($v_r =$ 3554
km s$^{-1}$) show that they have extended gas disks (\cite{Netal97}).
NGC~7673 is apparently face-on and fairly regular, with a mild
asymmetry on the western edge; NGC~7677 has a small outer extension
pointing towards NGC~7673.  The
\ion{H}{i} mass of NGC~7673 is $4 \times 10^{9}$ M$_{\odot}$, the dynamical
mass (rather uncertain due to the low inclination) 
inferred from this line profile is $3 \times 10^{10}$ M$_{\odot}$,
with the resulting fractional \ion{H}{i} mass to total mass of 2\%
(\cite{Petal01}).

The starburst activity is located within the inner half of the optical
radius, concentrated in large complexes, the `distinctive clumps' in
this galaxy (see Figure 4). 
The outer disk is relatively smooth and nearly circular,
but with ripples and wisps, likely remnants of a past interaction that
could have triggered the starburst. The overall 
distribution of optical light, however, is highly asymmetric, with a
rotational asymmetry value of $A=0.60$ (\cite{CBG00}). There is an
optical shell $\sim 21$~kpc west of the main body (\cite{Detal84}; 
\cite{HG99}); such features most frequently are seen around spheroidal 
systems,
but an optical shell is also present in the 
post-merger starburst galaxy
NGC~3310 (\cite{MvD96}).  NGC~7673 contains two strong non-thermal radio
emitting regions characteristic of Type~II~SN, 
indicating intense star formation occurred 
over the last $\sim 3 \times 10^{7}$ yr (\cite{CY90}).

Due to its starburst, NGC~7673 is bright in wavebands that are
sensitive to populations of young, massive stars, including the radio
(\cite{CY90}), far infrared (\cite{Ketal86}, \cite{Hetal89}), optical
emission lines (\cite{GHB89}, \cite{Getal97}), and ultraviolet
(\cite{BCH82}, \cite{Ketal93}).  Its integrated
SFR derived from the H$\alpha$ and far infrared luminosities is
$10-20$~$M_{\odot}$~yr$^{-1}$ for a Kroupa (2002b) parameterization of
the stellar initial mass function extending from 0.07 to 100~
M$_{\odot}$ (see \S 8.4).  Selected parameters of NGC~7673 are 
summarized in Table~1.

\begin{table*}
\caption[]{Parameters of NGC 7673}
\label{table1}
\begin{center}
\begin{tabular}{ccc}\hline
Parameter & Value & Reference \\\hline\hline
$\alpha$,$\delta$ (J2000) & 23:27:41.2, $+$23:35:21 & Falco et al. 1999\\
Redshift & 3408 km s$^{-1}$ & Huchra, Vogeley, \& Geller 1999\\ 
Distance & 49~Mpc & redshift and H$_{0} = 70$~km~s$^{-1}$~Mpc$^{-1}$\\
HI Mass & $4.09 \pm 0.03 \times 10^{-9}$ M$_{\odot}$ & Pisano et al. 2001\\
L$_{H\alpha}$ & $26.3 \times 10^{41}$ ergs s$^{-1}$ & Gallego et al. 1997\\ 
m$_{\it U}$ & 12.84 & De Vaucouleurs et al. 1991\\
m$_{\it B}$ & 12.87 & De Vaucouleurs et al. 1991\\
m$_{\it V}$ & 13.11 & Huchra 1977\\
M$_B$ & $-$20.7  & above with E(B$-$V)$=$ 0.04 \\
IRAS 12 microns & 0.14 $\pm$ 0.049 Jy & Soifer et al. 1989\\
IRAS 25 microns & 0.52 $\pm$ 0.042 Jy & Soifer et al. 1989\\
IRAS 60 microns & 4.98 $\pm$ 0.047 Jy & Soifer et al. 1989\\
IRAS 100 microns & 6.66 $\pm$ 0.321 Jy & Soifer et al. 1989\\
L$_{FIR}$ & $6.3 \times 10^{43}$~erg~s$^{-1}$ & IRAS 60 \& 100 micron flux after \cite{Hetal89}\\
\end{tabular}
\end{center}
\end{table*}

\section{Data and Observations}

The images analyzed in this paper 
were taken with the WFPC2\footnote{Based on observations made with the NASA/ESA Hubble 
Space Telescope, obtained from the data archive at the Space Telescope 
Science Institute. STScI is operated by the Association of Universities 
for Research in Astronomy, Inc. under NASA contract NAS 5-26555.}  and
reduced via the standard pipeline. In Cycle 5, two 600s exposures were taken
with the F555W filter, and two 400s exposures were taken with the
F814W filter. During this set of observations, the main body of the
galaxy was centered in the WF3 chip. In Cycle 6, three 800s exposures
were taken with the F255W filter.  Again, the main body of the galaxy
was centered in WF3, but with nearly a 90 degree offset 
in WFPC2 position angle from the
previous observations.

The images were combined with the IRAF\footnote{IRAF is provided by
the courtesy of the National Optical Observatories, which are operated
by the Association of the Universities for Research in Astronomy,
Inc., under cooperative agreement with the US National Science
Foundation.} task IMCOMBINE, with the rejection method CRREJECT to
remove cosmic rays. The images were divided by exposure time to yield
final images in DN s$^{-1}$. In the remainder of this paper we focus 
our attention on an investigation of the rich array of compact star 
clusters revealed in the WFPC2 images.

\subsection{Cluster Selection}

The IRAF packages DAOPHOT and APPHOT were used for the data analysis.
We identified objects with the DAOFIND task in all three images. For
the F555W and F814W images, we reran the photometry on each image
using the coordinate list from the other image. We then matched the
outputs and cleaned up any duplicates. From this list we matched the
transformed F255W coordinates, resulting in 5 lists of objects: F255W,
F555W, and F814W matched with each other, and F555W and F814W matched
with each other. We kept objects with errors less than 0.3 magnitudes
in the first set, and 0.15 magnitudes in the second set.  We excluded
objects fainter than 20.5 magnitudes in F255W, and objects redder than
1.4 in [$555-814$] in the two filter list (none appear in the three
filter list), which are most likely foreground stars.  With this color
criterion, we may be missing some dust obscured clusters, but without
additional filters, we cannot distinguish between a dusty cluster
and a foreground star. We also excluded 12 objects in the two filter
sample farther than 16 \arcsec ($\sim 4$~kpc) from the nuclear region
based on visual inspection. These objects were either obviously extended,
associated with large background galaxies, or had colors and 
magnitudes inconsistent with clusters, even
very old, very massive globulars. The two final sets contain 50 and 268 
objects, respectively.

While the data are sufficient to identify and perform photometry of 
clusters, they are
not optimal for systematically examining sizes and internal structures. 
Due to this lack of spatial resolution (1~pixel covers $\sim 25$~pc),
our clusters are not necessarily tightly bound, and may range from
relatively dense versions of OB associations to SSCs. Higher spatial
resolution imaging would allow a refinement of this sample.

Aperture photometry was performed using the IRAF task APPHOT with a 2
pixel aperture radius. Background subtraction was done with an annulus
of $5-$10 pixels. Charge transfer efficiency (CTE) 
corrections were applied according to Heyer
(2001). Gain corrections for all and aperture corrections for the
F255W photometry only were applied following Holtzmann et
al. (1995a \& b, hereafter H95a/b).  For the F555W filter, we
determined aperture corrections to the standard 5 pixel aperture using
three isolated, relatively bright objects available in the
WF3 field. For the F814W filters, two of these same objects were
bright enough for an accurate aperture correction measurement. We felt
that as uncertain as this is, it is still preferable to using the H95a
literature value. Aperture corrections determined in this way are
$-0.16$ mag for F555W and $-0.23$ mag for F814W.  
The comparison to the literature values for the
aperture corrections are $-0.16$ mag vs. $-0.14$ mag (H95a) for F555W,
and $-0.23$ mag vs.  $-0.18$ mag (H95a) for F814W. There are no
suitable objects in the F255W frame for determining aperture
corrections, thus, we use the values found in H95a.

We utilize the WFPC2 synthetic magnitude system, with photometric zero
points as listed in Table 9 of Holtzmann (1995b). The synthetic system
was defined so that Vega magnitudes match $UBVRI$ measurements in the
WFPC2 filter closest in wavelength to the $UBVRI$ filter. Thus the
F555W and F814W magnitudes are very close to V and I. We prefer to
stay in the WFPC2 synthetic system for comparison to the models, as
there is no satisfactory transformation from F255W magnitudes to U. We
present a histogram of transformed V and I values for comparison to
other studies in \S 7.

\subsection{Foreground Extinction and Reddening}

We have corrected our photometry for foreground Galactic reddening from  
Schlegel, Finkbiener, \& Davis (2000). At the position of NGC~7673 we 
adopt $A_{v} = 0.124$ and $E(B-V) = 0.04$. This amount of visual extinction 
translates into a neglible change of $-0.05$ mag in the [$555-814$] color, and 
$-0.13$ mag in [$255-555$]. 

\subsection{Completeness}

The completeness limit for our selection of nearly unresolved star
clusters was determined for the F555W filter by adding 10000
artificial stars at each 0.5 magnitude step between 20 and 27
magnitudes. The recovery fraction was calculated for each magnitude
step at a range of background levels.  At the highest background level
in this galaxy (500 DN), we have a 50\% recovery rate at 26th
magnitude in F555W. For 98 \% completeness at this background level,
the faint magnitude limit is 21.5.

\section{Models}
\begin{figure}
 \centering
 \includegraphics[width=8.5cm]{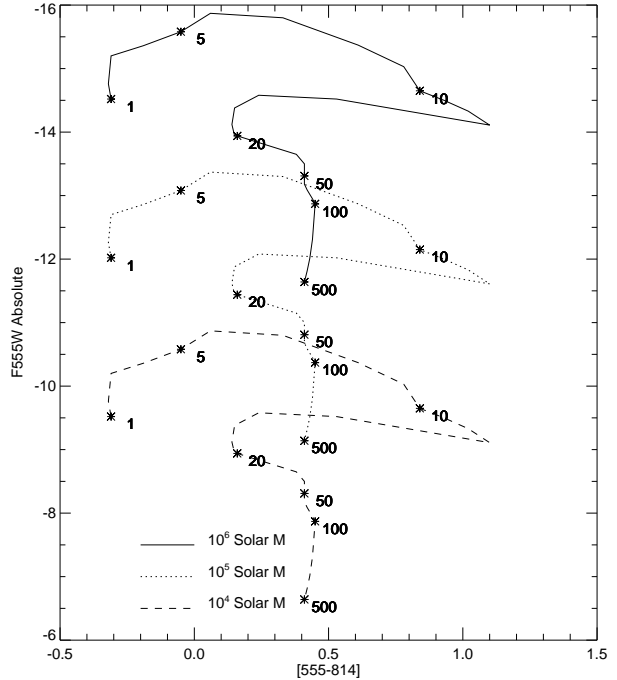}
 \caption{This figure presents Starburst99 (\cite{Letal99}) model tracks 
illustrating the effect of changing the mass of 
the star cluster. All model tracks are for an instantaneous burst. 
Asterisks mark the time at 1, 5, 10, 20, 50, 100, and 500~Myr. Note that
the only effect of a lower mass is to shift the tracks downward at 
constant color; stochastic effects are not included.}
 \label{massmodels}
\end{figure}

\begin{figure}
 \centering
 \includegraphics[width=8.5cm]{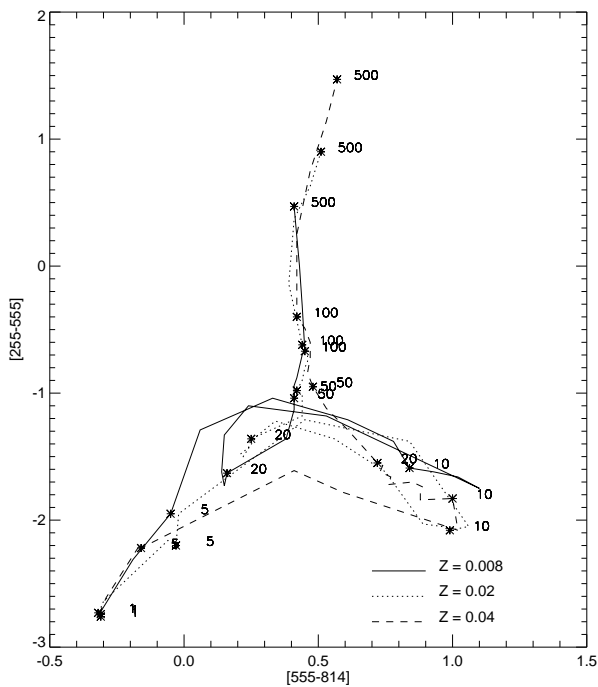}
 \caption{This figure presents Starburst99 (\cite{Letal99}) model tracks 
illustrating the 
effect of changing metallicity. All model tracks are for an 
instantaneous burst of $10^{6}$ M$_{\odot}$. Asterisks mark the time at 
1, 5, 10, 20, 50, 100, and 500~Myr.}
 \label{metmodels}
\end{figure}

\begin{figure}
 \centering
 \includegraphics[width=8.5cm]{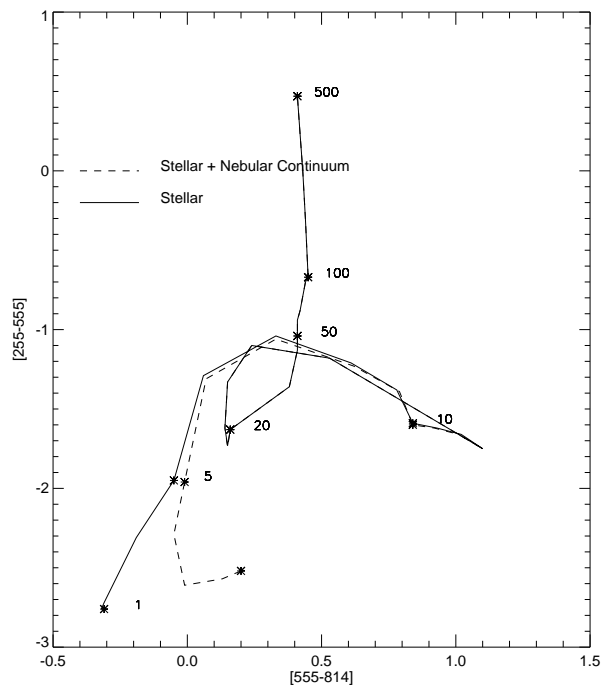}
 \caption{This figure presents Starburst99 (\cite{Letal99}) model tracks 
illustrating the 
difference between models with only stellar emission and those with 
stellar and nebular continuum emission for an instantaneous burst 
of $10^{6}$ M$_{\odot}$. Time steps are as for Figure \ref{metmodels}.}
 \label{nebmodels}
\end{figure}

We are using the Starburst99 evolutionary models (\cite{Letal99}) to
compare our colors and magnitudes. We calculate a magnitude for each
WFPC2 filter using standard response functions at the selected ages
and metallicities. Clusters are represented by an instantaneous burst
of $10^{6}$~M$_{\odot}$ with an upper mass limit of 100~M$_{\odot}$ and
a lower mass limit of 1~M$_{\odot}$. Varying the mass of the burst
affects the magnitude as shown in Figure \ref{massmodels}, but not the
colors. This means that our two-color comparisons are
mass-independent, insofar as stochastic effects are not
important in the real star clusters (see Lan\c con \& Mouhcine
2000). Varying the metallicity affects both the magnitudes and the
colors; this effect is shown for a two-color diagram in Figure
\ref{metmodels}. We do not account for stochastic effects, which
should introduce a spread in colors at cluster masses below $10^{5}$
M$_{\odot}$ (see \cite{Cetal02}).

For clarity, we have chosen to plot only a single metallicity track
for comparison with the NGC~7673 star clusters, $Z=0.008$, or just
under half Z$_{\odot}$. The abundance of this galaxy has been found to
be 12 + log O/H $\sim 8.6$ for the various clumps, slightly above that
of the LMC (\cite{DA82}), which is confirmed by a reanalysis of the
spectrophotometry reported by Gallagher, Hunter \& Bushouse (1989). 

\subsection{Nebular Emission}

The models we have chosen to use include only stellar
emission. Starburst99 also offers the opportunity to include nebular
continuum emission, calculated by assuming that every photon below 912
\AA~is converted into free-free and free-bound photons at longer
wavelengths (nebular case B). Although the model assumptions 
may not apply, for comparison we
show in Figure \ref{nebmodels} the variation in the two-color diagram
for star cluster models including 
stellar emission only vs. stellar + nebular continuum.  Here we
can see that the difference is significant only for ages $< 7$~Myr.

In young systems we also need to consider the 
influence of strong nebular {\it line}
emission, as well as the weaker nebular continuum.  Johnson et al. (1999)
investigated the effects of nebular emission lines on WFPC2 star cluster
colors. The main effect is due to strong emission lines of [OIII] and
H$\beta$ in the F555W filter. While the nebular continuum emission
makes the models redder in [$555-814$], they found the line emission
makes them bluer, and the difference can be large for ages $< 10$~Myr,
depending on the strengths of the emission lines relative to a 
cluster's stellar continuum (e.g., Olofsson 1989, Zackrisson et al. 2001).

Unfortunately, the issue is not as simple as deciding whether to
include nebular emission or not. For example, the nebular emission,
both continuum and line, may be subtracted as part of the background if it
extends smoothly beyond the cluster. So unless the nebular emission
characteristics are known, e.g., from high angular resolution spectra,
it is difficult to extract accurate ages from optical broad band
colors of very young star clusters. Nevertheless, we proceed with
fortitude, comparing our data with purely stellar continuum model tracks,
aware that nebular effects may be important. 

\begin{figure*}
\centering
 \caption{F555W mosaiced image. The lines are boundaries of the 
WFPC2 CCDs; the galaxy is mainly in the WF3 CCD with extensions into WF2
and WF4. The arrow indicates North (arrowhead) and East.}
 \label{bigpicture}
\end{figure*}

\begin{figure*}
\centering
 \caption{Left: A ground-based WIYN telescope 
R-band image showing the main clumps, which are labeled for 
reference following the designations of Duflot-Augarde \& Alloin (1982).
For both, North is up and East is to the left.
Right: A WIYN H$\alpha$ image showing that most of the 
clumps are strong H$\alpha$ sources. The exception is Clump D, which has
instead regions with H$\alpha$ emission to the NE and SE.}
 \label{RHa}
\end{figure*}

\begin{figure*}
\centering
 \caption{HST WFPC2 images, from left to right: F255W, F555W, F814W. 
 Orientation is as for Figure \ref{bigpicture}.}
 \label{3panel}
\end{figure*}

A rough estimate of the
kind of effect that could occur can be estimated from the 10~arcsec
circular aperture spectrum of Clump~A\footnote{Clumps are labeled following
the notation of Duflot-Augarde \& Alloin (1982)} presented in 
Gallagher et al. (1989). The combined
emission equivalent widths from the H$\beta$ line (corrected for
underlying absorption; as in Gallagher, Hunter, \& Bushouse 1989) and
the [OIII] $\lambda$5007$+ \lambda$4959 doublet is about 90\AA . This
then would produce an excess brightness of $\approx 0.07$~mag in the
F555W filter, and smaller offsets in the other two filters.
Of course this emission correction would be an underestimate if the
emission lines preferentially are associated with the individual star
clusters. We can derive a different correction for this case by assuming
all of the line emission is included in the photometry of the star
clusters in Table 6. In this case we find a substantial correction of
0.3~mag due to the emission lines in the F555W filter.

In summary, the figures subsequently found throughout this paper 
show only and always 
the model track for purely stellar emission from a $10^{6}$~M$_{\odot}$
cluster at a metallicity of $Z = 0.008$. 

\begin{figure*}
 \centering
 \caption{A color map made from the F555W and F814W images shows bluer objects
as darker, and redder as lighter. Orientation is as for Figure 
\ref{bigpicture}. Clump B is prominent in the top left 
(NE) corner of this image; Clumps C and F are also very blue. This image
also displays the complex small-scale dust structures associated with the
nuclear region (Clump A).}
 \label{colormap}
\end{figure*}

\begin{figure*}
\centering
 \caption{This F555W filter image of the central part of NGC~7673 shows the 
pair of linear dust structures emanating from a bright source, which may 
be the nucleus. Orientation is as for Figure \ref{bigpicture}.}
 \label{dustvee}
\end{figure*}

\begin{figure*}
\centering
 \caption{F555W image with the cluster sample overplotted as 
white dots and labeled with numbers corresponding to Table \ref{uvitab}. 
Orientation is as for \ref{bigpicture}.}
 \label{uviclus}
\end{figure*}
\section{Morphology}

Figure \ref{bigpicture} is a mosaiced F555W image with
a different scaling to illustrate lower surface brightness 
regions. Figure \ref{RHa} presents ground-based R and H$\alpha$ images 
obtained with the WIYN 3.5-m telescope; the reader is referred to HG 
for information about these data. 
Here they are presented to show the clump designations
by previous researchers, and for comparison with the WFPC2 images.
Figure \ref{3panel} shows the main body of NGC~7673 as observed 
by WFPC2 in the three filters: F255W, F555W, and F814W. 

From previous ground-based images, five main clumps have been 
identified (\cite{DA82}). These 
are designated as Clumps 'A--E'; here we refer to Clump A as the nuclear 
region, while Clumps B--F retain their original designations. 
Four of the clumps are strong H$\alpha$ sources, and three of the clumps
(B, C, and A/Nuclear) were observed in the Gallego et al. (1997) 
study of UCM galaxies. The
H$\alpha$ luminosity ranges from $2-8 \times 10^{8}$~L$_{\odot}$ in 
these regions. 
Both the nuclear region and Clump B are strong non-thermal radio sources,
with an inferred rate of 0.1 Type II SN~yr$^{-1}$ over the last 30~Myr
(\cite{CY90}).

The nuclear region (Clump A), has a pronounced bar shape, studded with
luminous star clusters. The spiral pattern is most noticeable on the 
north-eastern side of the galaxy, with an arm of $\sim$~2~kpc in length
emerging from the end of the bar to
play host to the highest surface brightness UV region in this system, 
Clump B. Small, stubby features, which suggest weak spiral armlets, extend 
from the nuclear region on the opposite side of the bar, which 
are also embedded with star formation regions.
Overall, the impression is of a disturbed, lopsided barred spiral,
without a distinct nucleus.

\subsection{Color Map \& Dust}

The color map shown in Figure \ref{colormap} was produced by dividing the 
F555W image by the F814W image, both in DN s$^{-1}$. This allows us to 
examine the distribution of dust and blue clusters in a qualitative way.
From this we can see that the clumps are prominently ``blue'' 
in the color map, excepting Clump~D. The dust is mainly concentrated
in and around the nuclear region, which has a conspicuous dust lane 
cutting across the center, and an overall patchy appearance. There are 
weak spiral arm features delineated by the blue clusters and dark 
dust lanes. 

Finally we note that in addition to the patchy dust extinction, faint but 
well-defined dust features extend in a ``V'' from a bright region in 
the nuclear clump (see Figure \ref{dustvee}). This peculiar structure may 
be the edges of a galactic chimney. The presence of a large scale outflow 
might be expected in a galaxy with a SFR as high as in NGC~7673.

\subsection{``Clump F'': No Clump, Just Cluster}

The region of the galaxy labeled ``Clump F'' appears distinctly 
non-clumplike in our WFPC2 images. Instead of an extended star formation 
region composed of many clusters
such as Clump B or Clump D, this region is dominated by a single object. 
The apparent F555W magnitude of this object is 19.5, and its inferred 
absolute magnitude is $-14.0$. This is extraordinarily bright compared
to young clusters in e.g. the Antennae (\cite{Wetal99}). In fact there is 
only one cluster in the Antennae sample which approaches this luminosity,
and even the dominant cluster in Hodge complex of NGC~6946 (which
at $-13.7$ contributes $17 \%$ of the light, 
\cite{Laretal02}) is less luminous than this. 

From Figure \ref{RHa}, we can see that
this object is embedded in a strong H$\alpha$ emitting region. Due to the low
resolution of the available H$\alpha$ data, we cannot deduce the amount of 
nebular emission coming from this cluster. It is possible the structure of the
emitting gas is complex, and the star 
cluster has blown out the gas in its immediate vicinity, allowing us to 
observe the unextincted stellar light. Closer inspection reveals that the
structure observed in the F255W and F555W is not the same, but the source
of this difference is unknown. Multiple stellar clusters, dust extinction,
or inhomogeneous nebular emission could be responsible.

\section{The Cluster Population} 

As stated in \S 2, we have two samples of star clusters: 
those detected in all three WFPC2 
filters, and those only detected in the F555W and F814W filters. We will
discuss each set separately, beginning with the three-filter set. For this
set of objects we have greater age discrimination, and we will use this  
information to investigate cluster age spreads both within and between 
the clumps. 

\subsection{F255W, F555W, \& F814W}

\begin{figure*}
 \centering
 \includegraphics[width=17cm]{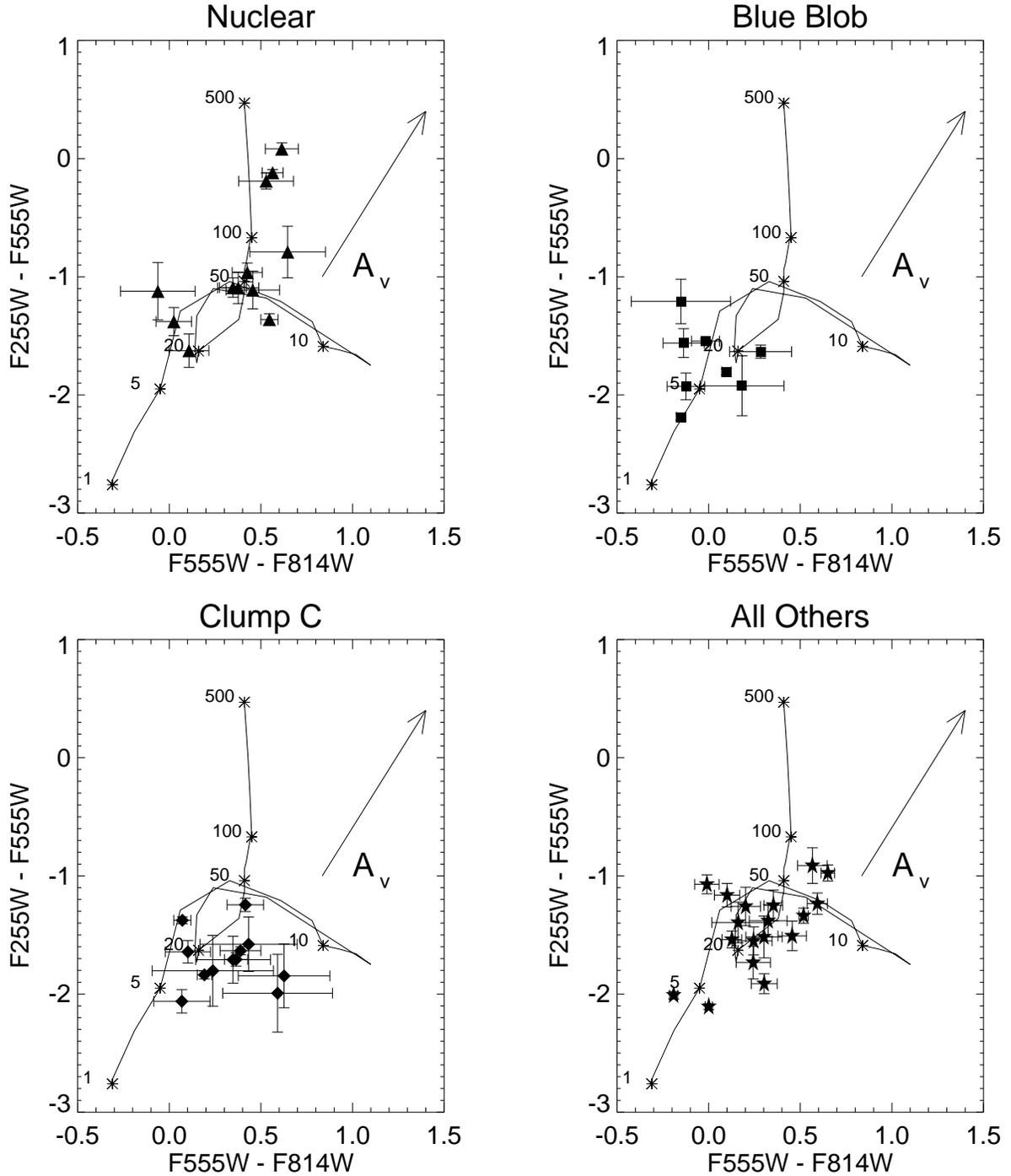}
 \caption{Two color plots for the clusters. 
Filled stars mark ages of 
1, 5, 10, 20, 50, 100, and 500~Myrs on a model track for purely stellar
emmission (see \S 4.1). Clockwise from top left: Nuclear,
the ``blue blob'' which is Clump B, all other clusters 
not associated with a particular clump, and Clump C.
The arrow indicates $A_{v}$=1. See discussion in the text.}
 \label{colorcolor}
\end{figure*}

\begin{table*}
\caption[]{Parameters for clusters detected in the F255W, F555W, \& F814W filters.}
\label{uvitab}
\begin{center}
\begin{tabular}{lllrrcl}
\hline
 Number & RA (J2000) & Dec (J2000) & [255-555]        & [555-814]       & F555W & Region \\\hline\hline
A-1....       & 23:27:41.19 & 23:35:18.92 & --1.09 $\pm$ 0.07 & 0.35  $\pm$ 0.08 & 19.80 $\pm$ 0.06 & Nuclear \\
A-2....	& 23:27:41.23 & 23:35:19.70 & --0.19 $\pm$ 0.15 & 0.53  $\pm$ 0.07 & 20.38 $\pm$ 0.05 & Nuclear \\ 
A-3....	& 23:27:41.19 & 23:35:19.48 & --1.36 $\pm$ 0.05 & 0.55  $\pm$ 0.05 & 19.53 $\pm$ 0.04 & Nuclear \\
A-4....	& 23:27:41.16 & 23:35:19.48 & --1.38 $\pm$ 0.10 & 0.02  $\pm$ 0.12 & 19.94 $\pm$ 0.08 & Nuclear \\
A-5....	& 23:27:41.14 & 23:35:19.39 & --1.63 $\pm$ 0.11 & 0.11  $\pm$ 0.14 & 19.91 $\pm$ 0.10 & Nuclear \\
A-6....	& 23:27:41.10 & 23:35:20.06 & --0.79 $\pm$ 0.21 & 0.65  $\pm$ 0.22 & 20.63 $\pm$ 0.15 & Nuclear \\
A-7.... 	& 23:27:41.08 & 23:35:20.18 & --1.12 $\pm$ 0.20 & --0.06 $\pm$ 0.24 & 20.66 $\pm$ 0.17 & Nuclear \\
A-8....	& 23:27:41.00 & 23:35:20.22 & --1.09 $\pm$ 0.11 & 0.38  $\pm$ 0.13 & 20.07 $\pm$ 0.09 & Nuclear \\
A-9....	& 23:27:41.07 & 23:35:20.64 & --0.96 $\pm$ 0.08 & 0.43  $\pm$ 0.08 & 19.97 $\pm$ 0.06 & Nuclear \\
A-10...	& 23:27:41.03 & 23:35:20.56 & --1.11 $\pm$ 0.15 & 0.46  $\pm$ 0.16 & 20.37 $\pm$ 0.11 & Nuclear \\
A-11...	&  23:27:41.04 & 23:35:20.93 & 0.08 $\pm$ 0.09 & 0.61  $\pm$ 0.05 & 19.07 $\pm$ 0.04 & Nuclear \\
A-12...	& 23:27:40.94 & 23:35:21.10 & --0.12 $\pm$ 0.06 & 0.56  $\pm$ 0.03 & 19.43 $\pm$ 0.02 & Nuclear \\
B-1....	& 23:27:41.70 & 23:35:29.86 & --1.80 $\pm$ 0.03 & 0.10  $\pm$ 0.03 & 18.88 $\pm$ 0.02 & Clump B \\
B-2....	& 23:27:41.65 & 23:35:29.57 & --1.54 $\pm$ 0.08 & --0.02 $\pm$ 0.02 & 20.40 $\pm$ 0.01 & Clump B \\
B-3....	& 23:27:41.63 & 23:35:29.90 & --1.56 $\pm$ 0.11 & --0.14 $\pm$ 0.12 & 20.44 $\pm$ 0.09 & Clump B \\
B-4....	& 23:27:41.70 & 23:35:30.51 & --1.92 $\pm$ 0.23 & 0.18  $\pm$ 0.25 & 21.10 $\pm$ 0.18 & Clump B \\
B-5....	& 23:27:41.66 & 23:35:30.35 & --1.21 $\pm$ 0.27 & --0.15 $\pm$ 0.19 & 21.01 $\pm$ 0.13 & Clump B \\
B-6....	& 23:27:41.72 & 23:35:30.79 & --1.63 $\pm$ 0.17 & 0.28  $\pm$ 0.06 & 21.33 $\pm$ 0.04 & Clump B \\
B-7....	& 23:27:41.62 & 23:35:30.68 & --1.93 $\pm$ 0.10 & --0.12 $\pm$ 0.11 & 20.22 $\pm$ 0.08 & Clump B \\
B-8....	& 23:27:41.63 & 23:35:31.17 & --2.12 $\pm$ 0.03 & --0.15 $\pm$ 0.03 & 18.98 $\pm$ 0.02 & Clump B \\
C-1....	& 23:27:41.41 & 23:35:25.20 & --1.64 $\pm$ 0.12 & 0.10  $\pm$ 0.09 & 21.29 $\pm$ 0.07 & Clump C \\
C--2....	& 23:27:41.33 & 23:35:25.08 & --1.80 $\pm$ 0.33 & 0.24  $\pm$ 0.30 & 22.22 $\pm$ 0.21 & Clump C \\
C-3....	& 23:27:41.37 & 23:35:25.73 & --1.58 $\pm$ 0.27 & 0.43  $\pm$ 0.23 & 21.80 $\pm$ 0.16 & Clump C \\
C-4....	& 23:27:41.33 & 23:35:25.64 & --1.99 $\pm$ 0.30 & 0.59  $\pm$ 0.33 & 22.26 $\pm$ 0.23 & Clump C \\
C-5....	& 23:27:41.36 & 23:35:25.97 & --1.71 $\pm$ 0.20 & 0.35  $\pm$ 0.20 & 21.55 $\pm$ 0.14 & Clump C \\
C-6....	& 23:27:41.29 & 23:35:25.55 & --1.85 $\pm$ 0.25 & 0.63  $\pm$ 0.27 & 21.90 $\pm$ 0.19 & Clump C \\
C-7....	& 23:27:41.28 & 23:35:25.91 & --1.63 $\pm$ 0.11 & 0.39  $\pm$ 0.03 & 21.24 $\pm$ 0.02 & Clump C \\
C-8....	& 23:27:41.34 & 23:35:26.41 & --1.71 $\pm$ 0.06 & 0.36  $\pm$ 0.06 & 20.14 $\pm$ 0.04 & Clump C \\
C-9....	& 23:27:41.37 & 23:35:27.25 & --1.38 $\pm$ 0.05 & 0.07  $\pm$ 0.03 & 19.83 $\pm$ 0.02 & Clump C \\
C-10...	& 23:27:41.34 & 23:35:27.11 & --1.84 $\pm$ 0.04 & 0.19  $\pm$ 0.03 & 19.89 $\pm$ 0.02 & Clump C \\
C-11...	& 23:27:41.27 & 23:35:27.11 & --2.06 $\pm$ 0.15 & 0.07  $\pm$ 0.10 & 22.32 $\pm$ 0.07 & Clump C \\
C-12...	& 23:27:41.36 & 23:35:27.78 & --1.25 $\pm$ 0.10 & 0.42  $\pm$ 0.06 & 21.01 $\pm$ 0.04 & Clump C \\
f-1....	& 23:27:42.00 & 23:35:16.92 & --1.34 $\pm$ 0.07 & 0.52  $\pm$ 0.03 & 20.71 $\pm$ 0.02 & field \\
f-2....	& 23:27:42.01 & 23:35:17.22 & --1.51 $\pm$ 0.13 & 0.46  $\pm$ 0.08 & 21.61 $\pm$ 0.05 & field \\ 
f-3....	& 23:27:41.29 & 23:35:16.00 & --1.55 $\pm$ 0.12 & 0.25  $\pm$ 0.10 & 21.73 $\pm$ 0.07 & field \\ 
f-4....	& 23:27:41.27 & 23:35:15.99 & --1.38 $\pm$ 0.14 & 0.32  $\pm$ 0.10 & 21.74 $\pm$ 0.07 & field \\
f-5....	& 23:27:41.04 & 23:35:17.95 & --1.34 $\pm$ 0.14 & --0.64 $\pm$ 0.10 & 21.65 $\pm$ 0.07 & field \\ 
f-6....	& 23:27:40.95 & 23:35:18.29 & --1.07 $\pm$ 0.08 & --0.01 $\pm$ 0.07 & 20.74 $\pm$ 0.05 & field \\
f-7....	& 23:27:40.84 & 23:35:20.08 & --1.16 $\pm$ 0.10 & 0.10  $\pm$ 0.07 & 21.20 $\pm$ 0.05 & field \\
f-8....	& 23:27:40.69 & 23:35:19.46 & --1.54 $\pm$ 0.07 & 0.13  $\pm$ 0.05 & 20.93 $\pm$ 0.04 & field \\
f-9....	& 23:27:41.25 & 23:35:22.99 & --0.98 $\pm$ 0.07 & 0.65  $\pm$ 0.04 & 20.41 $\pm$ 0.03 & field \\
f-10...	& 23:27:41.43 & 23:35:24.08 & --1.91 $\pm$ 0.09 & 0.30  $\pm$ 0.07 & 21.33 $\pm$ 0.05 & field \\
f-11...	& 23:27:41.29 & 23:35:23.87 & --1.23 $\pm$ 0.09 & 0.59  $\pm$ 0.06 & 20.96 $\pm$ 0.04 & field \\ 
f-12...	& 23:27:40.96 & 23:35:23.70 & --1.73 $\pm$ 0.14 & 0.24  $\pm$ 0.09 & 22.02 $\pm$ 0.07 & field \\
f-13...	& 23:27:41.00 & 23:35:24.65 & --1.91 $\pm$ 0.08 & 0.30  $\pm$ 0.07 & 21.33 $\pm$ 0.05 & field \\
f-14...	& 23:27:40.85 & 23:35:24.35 & --2.01 $\pm$ 0.03 & --0.19 $\pm$ 0.02 & 20.27 $\pm$ 0.02 & field \\
f-15...	& 23:27:40.66 & 23:35:23.71 & --1.26 $\pm$ 0.16 & 0.20  $\pm$ 0.08 & 21.75 $\pm$ 0.06 & field \\
f-16...	& 23:27:40.76 & 23:35:25.36 & --2.11 $\pm$ 0.03 & 0.00  $\pm$ 0.02 & 19.47 $\pm$ 0.01 & field \\
f-17...	& 23:27:40.71 & 23:35:25.63 & --1.39 $\pm$ 0.17 & 0.16  $\pm$ 0.14 & 21.52 $\pm$ 0.10 & field \\
f-18...	& 23:27:41.09 & 23:35:29.12 & --1.52 $\pm$ 0.18 & 0.30  $\pm$ 0.10 & 22.01 $\pm$ 0.07 & field \\\hline
\end{tabular}
\end{center}
\end{table*}

We have detected 50 objects with our selection criteria in the 
three filters: 12 in the nuclear region, 8 in Clump~B, 12 in Clump~C, and
18 others not associated with a particular clump. These are presented in
Table \ref{uvitab}, and overplotted as white dots with number labels 
on the F555W image 
in Figure \ref{uviclus}. Two-color diagrams for these objects are shown 
separated by region in Figure \ref{colorcolor}, with an overplotted
Starburst99 cluster model track for $Z=0.008$, with ages in~Myr. 
The reddening line is for $A_{v}=1.0$ using a Calzetti, 
Kinney, \& Storchi-Bergmann (1994) extinction law for starburst galaxies.
Mean [$255-555$] and [$555-814$] colors are presented in Table \ref{uvimeans}. 

From the color map in Figure 9, 
we would not expect all of the clusters to escape
the effects of dust extinction. However, we might expect to see only weak 
dust effects in our sample, because we are heavily biased in favor 
of clusters which
do not have significant reddening (for a discussion see \cite{Cetal00} and
\cite{C01}). 
One magnitude of visual extinction makes
a cluster approximately 2.4 magnitudes fainter in our F255W filter. Thus
we only expect to see the brightest, so the youngest and most massive, 
star clusters in this sample. At its most luminous, a $10^{6}$~M$_{\odot}$
cluster is predicted to have a F255W magnitude of $-17.9$ at an age of
3~Myr, or an apparent m(F255W)$\approx 15.3$ at the distance 
of NGC~7673. With our selection criteria we are sensitive to 3~Myr clusters 
down to $10^{4}$~M$_{\odot}$, but only $10^{5}$~M$_{\odot}$ clusters at 
20~Myr, due to rapid fading with age.

\begin{table*}
\caption[]{Mean [255-555] and [555-814] cluster colors for the three-filter 
selected sample.}
\label{uvimeans}
\begin{center}
\begin{tabular}{lccccc}
\hline
Region & Number of Members & Mean [255-555] (mag) & Stddev & Mean 
[555-814] (mag) & Stddev \\\hline
Nuclear      & 10   & $-0.90$ & 0.54 & 0.38  & 0.24 \\
Clump B      & 8    & $-1.72$ & 0.30 & 0.00  & 0.17 \\
Clump C      & 12   & $-1.70$ & 0.23 & 0.32  & 0.19 \\
No Region    & 18   & $-1.43$ & 0.33 & 0.23  & 0.30 \\\hline
\end{tabular}
\end{center}
\end{table*}

\subsubsection{Nuclear Region}

The nuclear region is 
unique as the only area significantly affected by dust in this galaxy.
This complicates our interpretation of the colors, because of the 
degeneracy between age and reddening. The reddening line 
can place a cluster at 1~Myr on the track anywhere from 2 to 100~Myr. 

The nuclear clusters span a wide range of [$255-555$] and [$555-814$] colors.
Three clusters (Numbers 4, 5, and 7) 
have blue [$555-814$] and [$255-555$] colors, and we suggest that 
these are young, $\sim$ 6~Myr clusters suffering from little extinction.
Four clusters (Numbers 1, 8, 11, and 12) concentrate near [$255-555$] 
$\sim -1.1$ and [$555-814$] $\sim 0.45$.
This places them on the model track at $30-50$~Myr. $A_{v}=0.5$ could make
20~Myr clusters appear here, and $A_{v}=0.8$ could cloak 5~Myr clusters 
and move them to the $30-50$~Myr position on the two-color diagram.
Three of the clusters (Numbers 2, 9, and 10) have [$255-555$] less than 
$-0.2$ and [$555-814$] $\sim 0.6$.
These could be 200~Myr with a small amount of reddening, or $7-50$~Myr with 
$A_{v}=1.0$.

Thus, we have 2 possible interpretations for the Clump A 
nuclear region star cluster population: 
(1) The clusters suffer little reddening 
($A_{v} < 0.2$ everywhere) in this region, and some of the clusters are 
young (t$_{age} \sim 6$~Myr), some are intermediate (30~Myr~$< 
t_{age} < 50$~Myr), and some are older (t$_{age} > 100$~Myr). Or, (2), 
there is significant and variable extinction ($A_{v} = 0.1-1.0$) 
in this region, and all of the clusters are $\sim 7$~Myr.
Additional information, such as high angular resolution
spectra (e.g., \cite{Glaze99}), is needed to resolve this issue.

\subsubsection{Clump B: The Blue Blob}

From the color map we find that this region has little, if any, 
reddening, which should simplify our
interpretation of the cluster colors. Looking at Figure \ref{colorcolor},
we can see immediately that almost all the objects are consistent with
ages $< 6$~Myr. The colors for
2 of the clusters (Numbers 3, and 5) are $0.1-0.2$ magnitudes too 
blue in [$555-814$], which could 
indicate contamination from nebular emission lines. We can easily mimic this 
effect by adding emission line fluxes 
to the F555W model magnitude. The clusters then overlap the 
models at ages $3-6$~Myr. Three of the clusters (Numbers 1, 4, and 6)
are consistent with ages $< 4$~Myr with a few tenths of a magnitude of visual 
extinction.

If we assume an age for the clusters, we can use the inferred absolute F555W
magnitude to estimate the masses. The clusters range from $-12$ to $-14.5$,
and assuming an age of 5~Myr, they span a range in mass of $5 \times 10^{4}$
to $5 \times 10^{5}$~M$_{\odot}$. 

\subsubsection{Clump C}

This region is complicated by a diffuse background in the F255W filter, 
indicating the presence of many faint F255W sources below our
detection limits. This is likely to be the case in 
all regions, however here there is a lack of very massive clusters as seen
in the nuclear region and Clump~B, which would prevent the detection of such
a background.

The Clump~C region does not match the models; the [$555-814$] colors are 
too red for their [$255-555$] colors (see Figure \ref{colorcolor}). 
However, Clump C is a strong H$\alpha$ source,
so we expect it to be young. A possible explanation for the colors
is that we are over-subtracting a strong nebular background.
This would make the F555W magnitude too faint. If we assume this is
the case, the [$555-814$] color gets bluer and the [$255-555$] gets redder,
moving them onto the model track around 6~Myr. The spread in 
F555W magnitudes is comparable to that of Clump B, but almost a 
magnitude fainter. For an age of 6~Myr, we estimate the star cluster masses
lie between $2 \times 10^{4}$ and $2 \times 10^{5}$ M$_{\odot}$. 

\subsubsection{Other F255W Sources: ``Clump F'' and the Central Cluster of 
Clump D}

What was designated as Clump F by previous researchers, is clearly dominated
by a single star cluster, number 16 of the ``no-region'' clusters in Table 
\ref{uvitab}. This object matches the model for a cluster with an age of 
$4-5$~Myr, with perhaps a small over-subtraction of nebular emission.
The sibling to this cluster, number 14 of the ``no-region'' 
objects, is also consistent with an age of $4-5$~Myr if we consider a 
contribution of 0.1 magnitudes in F555W from nebular line emission.
With this small nebular correction and the age assumption of 5~Myr, 
the clusters have masses of $3 \times 10^{5}$~M$_{\odot}$ and 
$1 \times 10^{5}$~M$_{\odot}$.

The central cluster of Clump D (Number 1) falls slightly below the model track 
for a cluster 
with an age of $9-13$~Myr, and its companion (Number 2) is off a bit farther, 
although along
the reddening vector. Assuming the photometry is reasonably accurate and the
models are not in error, these clusters can move to this area of the 
two-color plot
only by being very young, with ages of 1~Myr, 
and having more than 1 magnitude of visual 
extinction. 
However, they are unlikely to be younger
than 10~Myr, considering the lack of H$\alpha$ emission (see \S 4).
Also, we expect little to no reddening from the color map.
If we consider solar metallicity instead of Z$=0.008$, the model track moves 
blue-ward in [$255-555$],
eliminating any disagreement with the data. But it seems unlikely 
that the metallicity
in this region should be greater than that of the nuclear region. Rather, 
it seems more probable that Clump~D has a lower metallicity.

We also consider the possibility that the model tracks are in error. 
Allowing for the well-known  
uncertainty in red-supergiant stellar evolutionary tracks (\cite{LM95},
\cite{M97}, \cite{Oetal98}, \cite{Letal99}), especially at metallicities
less than solar,
it seems possible that the [$255-555$] color may be $\sim 0.2$ 
magnitudes too red. With all these possible factors, we prefer the 
explanation of a $9-13$~Myr cluster with little to no extinction.

An age of $9-13$~Myr also explains something noteworthy in the H$\alpha$
image. H$\alpha$ emission is not seen in the center of Clump~D, but
at the edges (NE and SE of the central cluster). Could this be evidence
for propagating star formation? Clusters are detected in these regions 
in the F555W and F814W filters (presented in \S 6.2.4), and while 
their colors are not 
remarkably blue, they are consistent with ages of $7-8$~Myr. This is 
interesting, although certainly not conclusive.


\begin{figure}
 \centering
 \includegraphics[width=8.5cm]{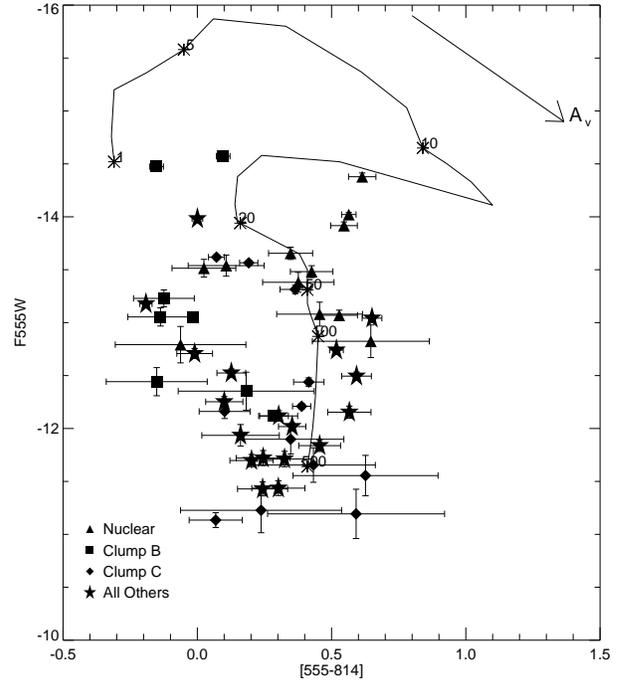}
 \caption{Magnitude-color plot for the 3-filter selected 
clusters shown
in Figures \ref{uviclus} and \ref{colorcolor}. The 12 nuclear clusters are
represented by triangles, the 8 Clump B clusters by squares, the 12 Clump C
clusters with diamonds, and the others not associated with a particular region
with stars. The model track is for an instantaneous burst of $10^{6}$ 
M$_{\odot}$ with Z=0.008. With an assumption of age, we can shift the model 
track vertically and derive a mass for the cluster based on its F555W
luminosity. The age assumption is most secure for the Clump B clusters 
($<$ 7~Myr); thus the brightest clusters in Clump B have masses 
around $5 \times 10^{5}$ M$_{\odot}$. }
 \label{onecm}
\end{figure}

\subsection{F555W \& F814W}

\begin{figure*}
 \centering
 \includegraphics[width=17cm]{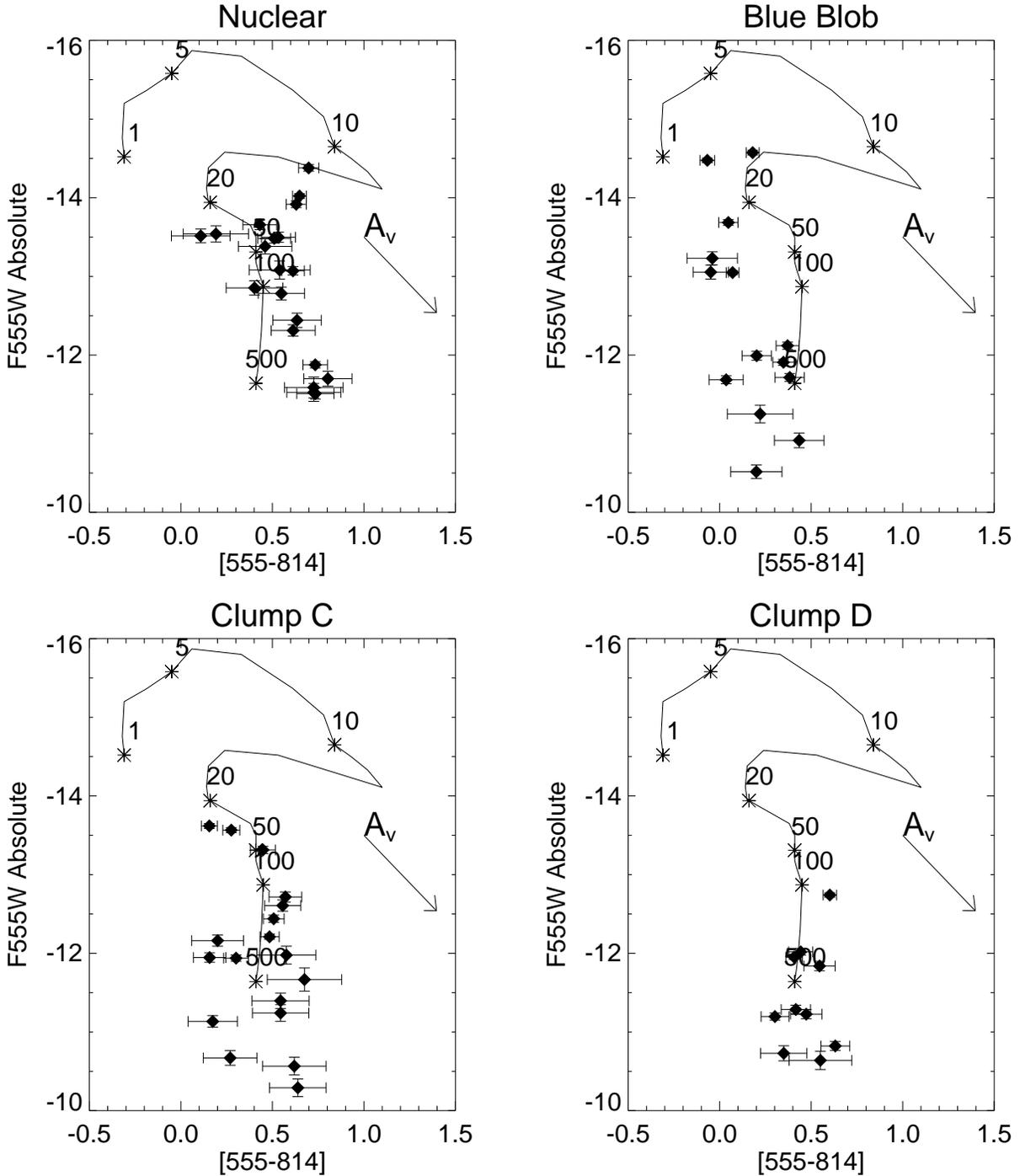}
 \caption{Magnitude-color plots for the clusters separated by region. The 
overplotted model track is for an instantaneous burst of $10^{6}$ 
M$_{\odot}$, with ages
of 1, 5, 10, 20, 50, 100, and 500~Myr marked with asterisks. 
A purely stellar emission model is used, i.e nebular contributions
are omitted (see \S 4.1). The track
shifts down by 2.5 magnitudes for $10^{5}$ M$_{\odot}$, with no change in
the F555W-F814W color (see Figure \ref{massmodels}).}
 \label{magcolor}
\end{figure*}

\begin{figure}
 \centering
 \includegraphics[width=8.5cm]{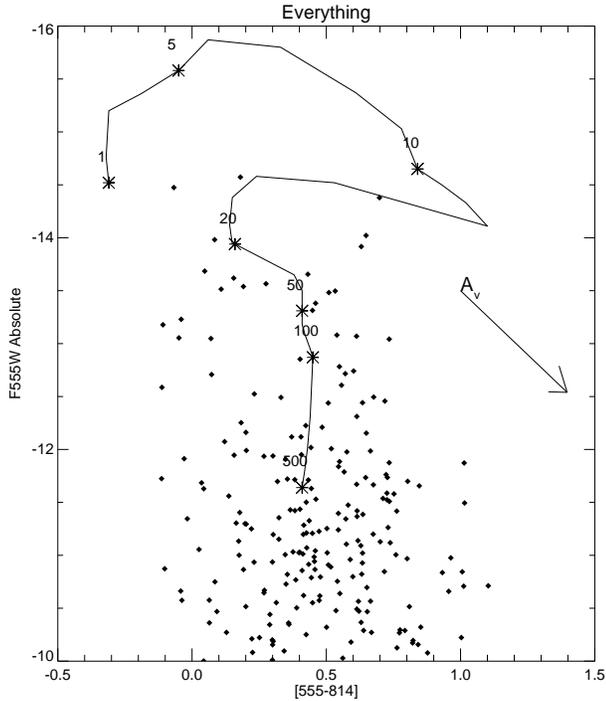}
 \caption{F555W magnitude and [$555-814$] magnitude color plot for the 268
two-filter selected clusters. The overplotted 
model track is for an instantaneous burst of $10^{6}$ M$_{\odot}$, with ages
of 1, 5, 10, 20, 50, 100, and 500~Myr marked with asterisks. The track
shifts down by 2.5 magnitudes for $10^{5}$ M$_{\odot}$, with no change in
the F555W-F814W color (see Figure \ref{massmodels}).}
 \label{everymagcolor}
\end{figure}

\begin{figure}
 \centering
 \includegraphics[width=8.5cm]{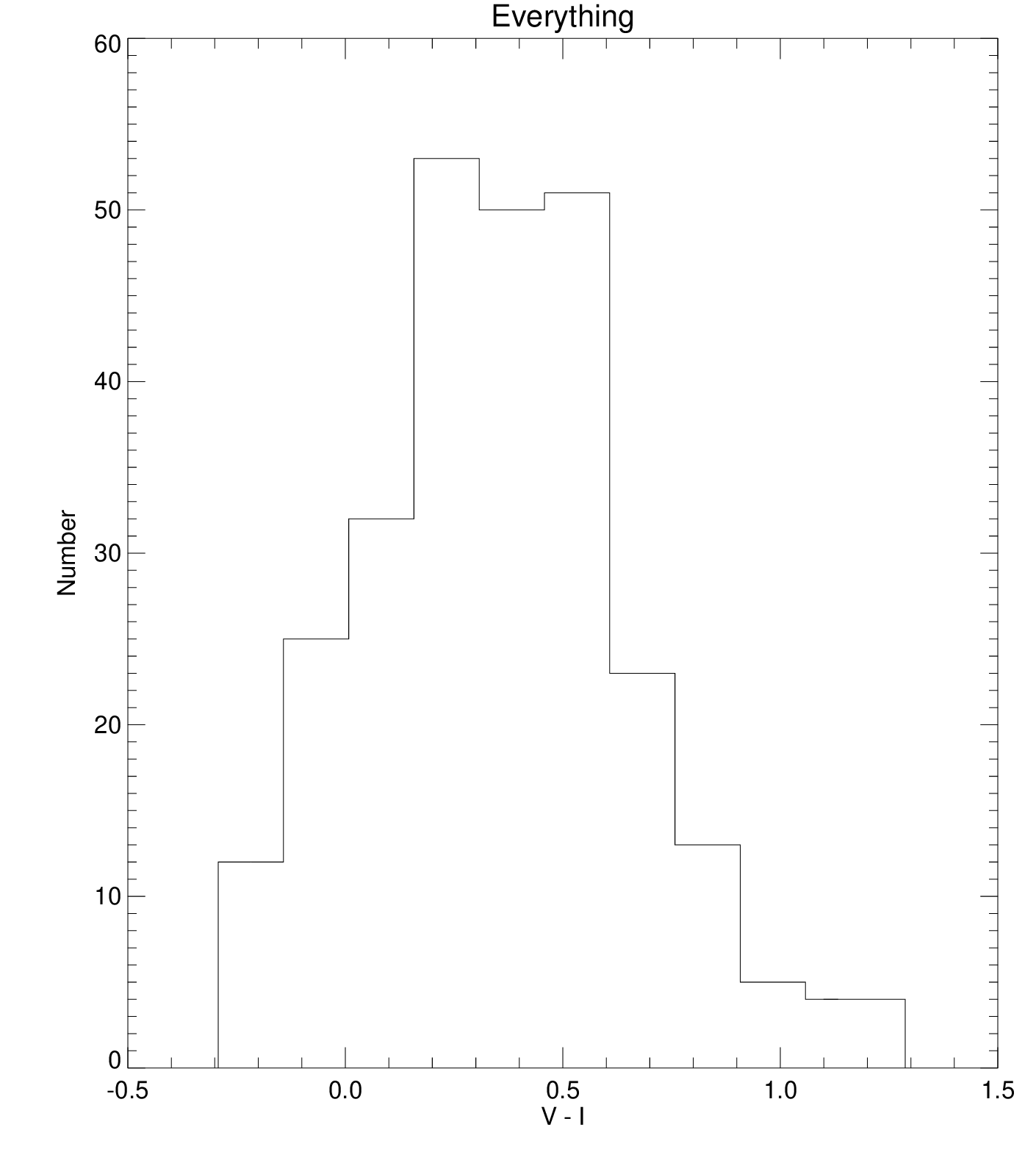}
 \caption{Histogram for the 268 clusters in the final 
F555W and F814W sample,
transformed to Johnson V and I. The mean color of this distribution is 0.42,
with a dispersion of 0.29.}
 \label{everyhist}
\end{figure}

\begin{figure*}
 \centering
 \caption{The two-filter cluster sample within the main galaxy body.
Orientation is as for Figure \ref{bigpicture}. From left of right: 
Blue clusters ($[555-814] < 0.4$), neutral ($0.4 < [555-814] < 0.7$), 
and red ($[555-814] > 0.7$.) The field of view for each panel is 
40~arcseconds, or 10~kpc at the distance of NGC~7673.}
 \label{rgb_.15}
\end{figure*}

Now we turn to the set of clusters selected in the F555W and F814W
filters, where we have detected 268 candidate star clusters with our 
selection criteria.
Splitting this population into regions as we did previously, we 
find 20 Clump A star clusters, 
14 clusters in Clump B, and 18 in Clump C. Here we 
also consider Clump D, which was too faint at 255, but in two filters
has 11 members. Thus, there are 200 candidate star clusters, mostly relatively
faint, which are not associated with a particular clump. 

The color magnitude plots for 
the clusters separated by region are shown in Figure \ref{magcolor}, and
the color magnitude diagram for the entire cluster sample is shown in Figure 
\ref{everymagcolor}. The 
clusters span a wide range of colors, with no apparent trend with 
luminosity. At the bright end are clusters in Clump B and the nuclear
region. The cluster population not associated with a particular region 
starts to appear about two magnitudes fainter than these. This means that the
most massive clusters are born in the clumps, although cluster formation
is occurring throughout the galaxy.

For comparison with studies of other star-forming galaxies, 
we convert our [$555-814$] colors to
standard Johnson V and I using the transformations found in H95b. 
The histogram of cluster colors is shown in Figure
\ref{everyhist}, using a bin size of 0.15 magnitudes. 
Mean $V-I$ colors are shown in Table \ref{vimeans}, along with the mean
[$555-814$] colors. 

The mean color of this distribution, 0.42, is blue, and comparable
to the young cluster population in the Antennae (0.4, age $< 30$~Myr,
 \cite{Wetal99}), the 
blue cluster population of NGC~7252 (0.4, age $\sim 30$~Myr, 
\cite{Metal97}), and the 
cluster population of the 4~Myr starburst in NGC~1741 (0.4, \cite{Jetal99}). 
This suggests a similarly young age for the cluster population presented
here. It is clearly younger than the cluster population of NGC~3921 (0.59, 
age $\sim 250$~Myr, \cite{Setal96}).

\begin{table*}
\caption[]{Mean V--I and 555--814 cluster colors for the two-filter selected sample.}
\label{vimeans}
\begin{center}
\begin{tabular}{lcccc}
\hline
Region & Number of Members & Mean [V--I] (mag) & Mean [555--814] (mag) & Stddev (mag)\\\hline
Nuclear      & 20   & 0.47 & 0.56 & 0.18 \\
Clump B      & 14   & 0.07 & 0.17 & 0.17 \\
Clump C      & 18   & 0.33 & 0.43 & 0.18 \\
Clump D      & 11   & 0.40 & 0.49 & 0.13 \\\hline
All Clusters & 268  & 0.42 & 0.52 & 0.29 \\\hline
\end{tabular}
\end{center}
\end{table*}

\subsubsection{Nuclear}

The mean color of this region is redder than the cluster population
as a whole. Looking only at the color map, we would suspect this
to be the case due to dust extinction. From Figure \ref{magcolor},
we can see this is indeed what is happening. There is a 
pronounced color-luminosity 
trend, where the faintest clusters are the reddest ($[555-814]\sim 0.65$). 
There are two possible
causes for a color-luminosity trend, age and reddening. We can rule out 
an aging effect since the [$555-814$] color of a cluster is nearly constant 
as it ages. Therefore, the observed trend must be due to dust extinction.
There are three important exceptions to this trend; the three brightest 
clusters have red colors ($[555-814] > 0.5$). Either these are old and
very massive clusters, or they are very young and have yet to 
destroy their dusty natal cocoons. This can only be determined from
spectroscopy, or, perhaps from high angular resolution 
images in emission lines that would reveal connections
to the surrounding ionized gas.

\subsubsection{Clump B: The Blue Blob}

The color spread in this region is small, approximately 0.2 magnitudes in 
[$555-814$]. However, there is a tendency for less luminous clusters to be 
redder, suggesting the effects of a small amount of reddening, requiring 
a correction for $\sim$~$A_{v}=0.3$ magnitudes. If we make this small
correction to the faint clusters, the color of this region becomes 
nearly zero in [$555-814$], and the color spread becomes even smaller, 
$\sim 0.15$ mag. Now 
we can say something about the formation timescale of this region. 
From a dynamical point of view, we would not expect a 
region of this size, radius $\sim 0.4$~kpc, to have an age spread less 
than 3~Myr (using $v = 150$~km~s$^{-1}$). But for this small range in color,
the age spread cannot be more than 3~Myr, could be smaller, and
is only compatible with the models if the entire region is younger 
than 6~Myr. Thus, we conclude that the clusters in Clump~B are less than
6~Myr with age spread $\leq 3$~Myr. With this age,  
cluster masses range from $5 \times 10^{5}$ to $5 \times 10^{3}$ M$_{\odot}$. 

\subsubsection{Clump C}

The mean cluster color in this region is redder than that of 
Clump~B, closer to that of 
Clump~D, but with a larger color spread. Only for the brightest clusters
does there appear to be a trend of color with luminosity. The mean color
of this region suggests an age less than 50~Myr, although the colors
of the three-filter sample suggests that at least some of the clusters 
are less than 6~Myr. As we saw for the three-filter sample, there
may be complicating nebular effects in this region.

\subsubsection{Clump D}

The clusters associated with Clump~D all have roughly the same color,
but a range of luminosities. This could mean that they have the 
same age and different masses, or
a small range in mass, and ages $< 30$~Myr with a spread. 
With a radius of approximately 200~pc, an age 
spread greater than 1.3~Myr is expected.
This region is significantly fainter 
than the other star forming regions in this galaxy, with the brightest 
member at only 20.7 magnitude at F555W, or $-12.7$ 
in absolute magnitude. For the derived age
of $9-13$~Myr (as discussed in the previous subsection), 
we find a mass of $2 \times 10^{5}$~M$_{\odot}$.

\section{Clump Properties}

To investigate the clumps in more detail, we have performed photometry with
POLYPHOT for the nuclear region, and APPHOT with radii of 20, 22, and 
19 pixels for Clumps~B-D. The results are shown in Table \ref{clumps}.
For the [$255-555$] numbers, we have used the three-filter selected objects,
while for the [$555-814$] set, we have used the two-filter selected set of 
objects. 

We can immediately see that the nuclear region is the reddest in both
colors. Regardless of the age of the latest burst in this region, we expect
this to be the case from the color map. Clump~B is bluest in [$555-814$]
color, and, perhaps surprisingly, Clump~C is bluest in [$255-555$] color. 

We can use these measurements to estimate the efficiency of cluster
formation by calculating the flux from clusters vs. the
total flux from the region; these results are shown in Table \ref{frac}.
Measurements of the fraction of light from clusters 
at 2200 \AA~in a sample of starburst galaxies found an average of 
$\sim 20 \%$ (\cite{Metal95}). The numbers here are similar, and 
range from $13 \%$ near $I$ for the 
nuclear region, up to $33 \%$ for the mid-UV flux coming from clusters in 
Clump B. At F555W, the total flux in clusters compared to the 
total galaxy flux is $7 \%$.

The F255W filter gives us information about the stellar emission
{\it only},
and primarily about the young, massive stars which dominate at this 
wavelength. One consequence of this
is that a measurement of the fraction of the 
F255W flux coming from the clusters compared to the overall flux gives 
a direct measurement of the massive star formation occurring below our cluster 
detection limits. In contrast, the F555W and F814W filters tell us about the 
combination of stellar and nebular emission. 

Clump~B has $33 \%$ of its UV emission coming from only 8 clusters, but 
this number drops to $22 \%$ of visual light coming from 14 clusters, and
$20 \%$ at F814W. Clump~C has $22 \%$ of its UV emission from 12 
clusters, but this number rises to $26 \%$ at F555W and F814W with a 
total of 18 clusters. We interpret this as evidence for an underlying 
stellar population in Clump~B, which is absent in Clump~C. Although 
Clump~B is a strong H${\alpha}$
source, we can rule out nebular emission not associated with the clusters,
as this would make the [$255-555$] and [$555-814$] colors too blue and
too red, respectively. 

\begin{table*}
\caption[]{Photometric properties of the Clumps}
\label{clumps}
\begin{center}
\begin{tabular}{lccccc}
\hline
Region &    F255W & F555W & F814W & [255--555] & [555--814] \\\hline
Nuclear   & 14.22 & 14.84 & 14.23 & $-0.62$   &  0.61   \\
Clump B   & 14.56 & 15.81 & 15.64 & $-1.25$   &  0.17   \\
Clump C   & 14.91 & 16.48 & 16.07 & $-1.57$   &  0.41   \\
Clump D   & 16.33 & 17.31 & 16.89 & $-0.98$   &  0.42   \\\hline
\end{tabular}
\end{center}
\end{table*}

\begin{table*}
\caption[]{Fraction of Flux from Clusters vs. Total Flux}
\label{frac}
\begin{center}
\begin{tabular}{lccccc}
\hline
Region & F255W & F555W & F814W \\\hline
Nuclear      & 0.16 & 0.14 & 0.13 \\
Clump B      & 0.33 & 0.22 & 0.20 \\
Clump C      & 0.22 & 0.26 & 0.26 \\
Clump D      & .... & 0.16 & 0.17 \\\hline
\end{tabular}
\end{center}
\end{table*}

\section{Discussion}

\subsection{Clump Summary}

We have shown that the clumps are composed of many bright super 
star clusters, and that many of these are less than 6~Myr.
This conclusion remains, despite the difficulties in deriving
ages below 7~Myr due to nebular emission and unknown geometry.
The nuclear region contains a bar of star clusters, some of 
which are certainly 
young, and others which may be old, but are likely also young and 
suffering from dust extinction. 

Clump~B is a luminous, high surface brightness, and very young region,
which has some evidence for an older, underlying stellar population.
The young star clusters have ages less than 6~Myr, with masses ranging from 
$5 \times 10^{4}$~M$_{\odot}$ to $5 \times 10^{5}$~M$_{\odot}$. For these 
8 clusters, we have a combined mass of approximately $8 \times 
10^{5}$~M$_{\odot}$.
 
We have shown in \S 7 that $33 \%$ of the flux at F255W is coming from 
these clusters, thus $2/3$ of the massive star formation is below our
detection limits. We can use this to estimate the SFR in this region 
over the last 6~Myr, if we assume a constant M/L. 
If in the $1/3$ of the light that we observe we have $8 \times 
10^{5}$~M$_{\odot}$, and we are missing $\sim 2/3$ of the massive 
stellar flux, this implies that
the total mass formed in this region is $\sim 2.4 \times 
10^{6}$~M$_{\odot}$. With a radius of 0.4~kpc, Clump~B then has a 
SF intensity
of 0.8 M$_{\odot}$ yr$^{-1}$ kpc$^{2}$. This incredible SF density puts it
in the league with the most vigorously star-forming galaxies in the universe. 
Lanzetta et al. (2002) found objects at z $\ge 3$ with similar SF intensities
over significantly larger areas than that of Clump~B. However, Clump~B
still stands out as a large and intense extra-nuclear star forming site
that can provide insights into the consequences of intense
star formation on relatively large spatial scales.

Clump~C is undoubtably young judging from its blue overall [$255-555$] color 
and those of the composing clusters. It may have embedded clusters 
judging from 
the combination of H${\alpha}$ emission and the anomalous cluster colors 
in our two-color diagrams.

Clump~D is composed of fainter clusters, pointing to an older age and
possibly less massive clusters. 
The derived age of the brightest cluster
is between 9 and 13~Myr, and there are regions to the NE and SE with H$\alpha$
emission, characteristic of young star formation regions, i.e those with ages
of less than 10~Myr. We suggest that this is evidence of
propagating star formation, which should be followed up with high spatial
resolution spectroscopy. 

Clump~F is not a clump but an \ion{H}{ii} region dominated by a 
single cluster with an age of $4-5$~Myr. This object is extremely bright, 
and the disparity between the F255W and F555W structures indicates 
smaller scale structure than we can resolve associated with dust or multiple
ages.

The clumps in NGC~7673 appear to represent a step beyond normal 
OB associations in terms of their hierarchy of compact, gravitationally bound 
objects. In a simple OB association, the rare massive stars, some of 
which are either binary or multiple systems, define the upper end of 
the hierarchy. They contain only a modest fraction of the stellar mass, and 
are surrounded by more numerous, less massive single and multiple stars, 
which contain most of the mass. 

In starburst clumps, compact star clusters 
define the top of the compact, bound mass distribution, 
and contain a significant fraction 
(more than 16\%) of the recently formed stars (see \S7). These in turn 
are embedded in more diffuse fields of massive stars, which resemble 
rich examples of OB associations. Since high redshift galaxies also show 
signatures of starburst clumps, this mode of star formation, while rare 
in the current epoch, may have played an important role in building 
galaxies. As one of the nearest optically accessible clumpy starburst galaxies, 
NGC~7673 provides an excellent opportunity for further exploration of 
intense, large scale star formation.

\subsection{Starburst Picture: Age of the Current Burst}

HG (1999) discussed the starburst trigger in detail, concluding that 
either the burst was triggered by an interaction with its companion, 
NGC~7677, or by the consumption of a small galaxy less than $10 \%$ 
of its mass in a minor merger. Neither of these scenarios can be
ruled out, but in both cases the model must be able to account
for the following: 

1. The H I map of this pair shows a few 
appendages to the disk of NGC~7673, and a small extension from NGC~7677
pointing toward its companion, but otherwise is surprisingly regular.

2. NGC~7673 has an outer optical shell. This type of feature is usually 
associated with merger candidate E, S0, and only a few Sa galaxies
(\cite{SS88}), but has also been found around the starburst galaxy
NGC~3310, which is thought to be triggered by a minor merger(\cite{MvD96}). 
Theoretical work indicates, however, that interaction
scenarios other than mergers can produce arc-like features outer disks,
like the one seen in NGC~7673 (\cite{HQ89}, \cite{HS92}, \cite{Hetal93}). 

3. There is a difference in timescale between the formation of this outer disk
ripple and the very young starburst occurring in the inner disk. The outer 
disk ``ripple'' is characteristic of the late phases of an interaction, 
yet the starburst remains quite active. For example, if the starburst 
is associated with a close passage of NGC~7673's disturbed neighbor, 
NGC~7677 at a projected distance of 95~kpc, then the triggering 
event probably took place several hundred Myr in the past. 

That O stars are present through out the galaxy is clear from the 
prominence of \ion{H}{ii} regions, indicating star formation within 
the last 10~Myr (see HG). Duflot-Augarde \& Alloin 
(1982) also found evidence for an underlying older stellar population within
the nuclear region from a weak 400 nm continuum break and Balmer
absorption underlying the strong emission lines. However, the digital 
spectrum of the nuclear region taken by Gallagher et al. (1989) 
suggests that the 4000~\AA~``break'' is probably due to a reddened younger 
population. Duflot-Augarde \& Alloin also found
weak evidence from the G band for the presence of an evolved stellar 
component in Clump B. This is supported by the fraction of flux at F555W
from cluster as opposed to F255W. So the starburst history in this galaxy
is complex.

Our two-color diagrams are useful in age-dating the brightest regions,
although the rapid fading and sensitivity to dust of the F255W flux
means we have only a small number of the entire cluster sample for
this type of investigation. With this we find evidence for very 
young, $< 6$~Myr star clusters in Clump B, Clump C, and the nuclear region. 

The color magnitude diagram for the entire cluster population
shows a broad range in [$555-814$] and V$-$I color, spanning the allowed
model color range fully. It is difficult to constrain ages with only
this color, although we do note that the mean is comparable to other
starburst systems with burst ages less than 30~Myr (such as the Antennae
and NGC~7252), and one as young as 4~Myr (NGC~1741).

An overview of the distribution of cluster ages is illustrated in 
Figure \ref{rgb_.15}. This shows the bluest star clusters in [$555-814$],  
whose intrinsic colors imply ages of $< 20$~Myr, 
are concentrated in the blue blob, Clump B, with some presence 
in Clumps C and F. These objects avoid the inner parts of the starburst, 
yet Clump A also displays strong H$\alpha$ emission (see Figure \ref{RHa}). 
The possible conclusions are that the H-ionizing stars in Clump A are 
distributed in the field rather than in compact star clusters, or dust 
is an important factor. 

The middle panel of Figure \ref{rgb_.15} shows star clusters with colors 
consistent with ages of about $20-30$~Myr up to 1 Gyr if they are unreddened 
or moderately reddened younger star clusters. Note these are absent in 
Clump B, where we see little evidence for dust; clump B apparently is a 
young feature. The combination of `blue' and `neutral' star clusters 
delineate the main features of NGC~7673. The strong luminosity bias 
that favors observation of younger star clusters, the coincidence between 
the locations of these clusters and H$\alpha$ emission, and the results from 
the F255W filter observations lend additional support to the 
view that the clumps in NGC~7673 are relatively young, with ages of 
$\le$50~Myr. 

In this interpretation the upper age bound is set by Clump D, 
which has weak H$\alpha$ emission, faint star clusters with similar 
colors, and comparatively red global colors (Table 5). The third panel 
of Figure \ref{rgb_.15} shows cluster candidates which either are 
highly reddened and young, luminous objects, or have colors of star 
clusters with ages of 1~Gyr or more. We are encouraged that this poorly 
defined sample of redder objects is not associated with the 
main clumps. Whatever the 
red objects are (possibly a mixture of background galaxies, clusters, 
and objects with large color errors), they do not seem to be associated with 
the ongoing starburst.

Using the clump colors in Table 4, we can estimate ages, which will lie 
between the values predicted by instantaneous and continuous SFR models. 
We again use the Starburst99 models. Clump B then must predate the red 
supergiant flash, and is younger than about $8-10$~Myr. Clump C is 
$10-15$~Myr old, while Clump D is $15-50$~Myr in age. These ages refer to the 
mean stellar population age. They are not corrected for any internal 
reddening, and including these would reduce the estimated ages.

\subsection{Evolution of the Starburst Clumps}

Huge regions of active star formation, such as those seen in NGC~7673,
presumably reflect the presence of supergiant complexes of
gravitationally bound interstellar gas clouds. These can form when a 
galactic disk becomes Jeans unstable. The Jeans mass for a rotating 
gaseous disk
scales as $M_J \propto \overline{\sigma_g}^{2}/\mu_g \propto \mu_g^3 Q_g/
\kappa^3$, where $\mu_g$ is the gas disk's surface density,
$\overline{\sigma_g}$ the velocity dispersion, $\kappa$ the orbital
epicyclic frequency, and $Q_g \sim 1$ the Toomre disk stability
parameter (\cite{EKT93}, \cite{N99}).  Galaxy interactions favor the
formation of super cloud complexes in two ways: (1) They can increase
the gas velocity dispersion $\overline{\sigma_g}^{2}$ (\cite{EKT93}). (2) They tend to drive gas inwards, either as a direct
result of the interaction, or indirectly through the presence of
tidally-induced bars (e.g., \cite{N87}, \cite{BH96}).  We might expect
to find starburst clumps in disturbed, gas-rich disk galaxies, such as
NGC~7673.

Our study of NGC~7673 suggests it contains two classes of clumps. Clump A, 
the nuclear region, lies near the middle of an offset stellar bar. It 
is a special place, where dynamical friction and dissipation can pile 
up material that has been transported inwards, possibly leading to 
the formation of a bulge (\cite{N99}). The other clumps are located 
beyond the bar. A key question is whether the outer clumps can  
survive for sufficiently long to carry their material inwards to 
make a bulge upon their disruption near the center of the galaxy, as 
hypothesized by Noguchi (1999).

We take Clump~D, with M$_B \approx -$16.3 as an example. Using the 
Starburst99 
models and a Kroupa stellar IMF (2002), we estimate that 
(M/L)$_V \approx$0.03 for a 
constant SFR age of 30~Myr. Clump~D then has $M \ge 10^7$M$_{\odot}$ or 
about 0.25\% of the NGC~7673 dynamical mass; 
this mass is a lower limit as we have not included any gas. Taking the 
surface density of Clump~D to be 5 times that of the mean disk, the Noguchi 
(1999) model then  predicts the dynamical friction time scale for Clump~D to 
reach the center of the galaxy is a few orbital periods, or $>$ 100~Myr.

While the non-nuclear clumps have a range in ages, with Clump~B 
being the youngest, none shows compelling evidence for star clusters 
with ages of $\ge$100~Myr. This in part reflects the shallow nature of 
this initial survey of NGC~7673, which means we will miss any but the 
most massive older star clusters (see \S 6). However, we have seen that 
the integrated V$-$I colors of the outer clumps also imply ages of 
$<$100~Myr. 
Furthermore, we do not see evidence for inward migration of older 
clumps; both the young Clump B and older Clump D are located at 
similar radii. Our data therefore do not reveal the presence of long-lived 
clumps in NGC~7673, and so possibly, as also discussed by Noguchi (1999), 
this is a case where the clumps are marginally bound and less durable than 
in protogalaxy where more of the mass is in the form of gas. 

However, we also recognize that we are observing NGC~7673 well after whatever 
interaction triggered its starburst. Perhaps earlier  
generations of clumps were more 
robust and already have accreted into the central region of the galaxy?
After all, more than 100~Myr are likely to have passed since the starburst 
was triggered, an adequate amount of time for dynamical friction to act. 
In this case we would expect Clump A to contain an unusually wide spread 
in star cluster ages from the previously dissolved clumps. Unfortunately, 
the interpretation of star cluster colors in the nuclear region is 
complicated by a combination of crowding and dust. Despite these 
complications, Figure \ref{rgb_.15} shows that the redder clusters are 
not overly concentrated in Clump A, as might be expected if a few large 
clumps had dissolved in the not too distant past.

\subsection{Final Fate of NGC 7673}

What will ultimately happen to NGC~7673? In their study of LCBGs, 
Pisano et al. (2001) calculate $\tau$$_{gas}$, the \ion{H}{ii} 
mass divided by the star formation rate, to give
a simple estimate of the gas depletion timescale. Using their H$\alpha$ 
luminosity as the SFR indicator, they find SFR~=~$23.5$~M$_{\odot}$~$yr^{-1}$ 
with the Kenicutt (1983) 
calibration, yielding $\tau$$_{gas}$~=~0.18 Gyr. If we use a more modern 
SFR calibration from combining a Kroupa parameterization of the IMF 
with Starburst99 Lyman continuum luminosity predictions, then we find 
SFR~$\approx 12$~M$_{\odot}$~yr$^{-1}$ from the Pisano et al. L(H$\alpha$).
Using the L$_{FIR}$ from Table 1, we get a similar result, and conclude 
that the current SFR is in the $10-20$~M$_{\odot}$~yr$^{-1}$ range. 
Adopting M$_{gas} = 1.3$~M$_{HI}$ to allow for the presence of helium, 
we get an only slightly more conservative $\tau_{gas} \approx$0.5~Gyr. 
This is short, and if
NGC~7673 continues to form stars at the current rate, it will consume 
the rest of the available gas within the next 0.5~Gyr. 

If it consumes all of its gas during its  
fast and furious starburst phase,  then NGC~7673 will 
become a nearly gas-free disk
galaxy, probably a small Sa or S0 system. 
If, on the other hand, the starburst is beginning its decline, 
this could allow a substantial fraction of 
the remaining $4 \times 10^{9}$~M$_{\odot}$ of \ion{H}{i} to form stars 
more sedately. The resulting system 
would then appear to be a late-type galaxy. \ion{H}{i} masses 
of $\sim$10$^9$~M$_{\odot}$ are typical of small star-forming disk galaxies, 
and since NGC~7673 has been collisionally perturbed, we believe it would 
most likely have a somewhat thickened disk, such as those found in Magellanic 
spirals. 

\section{Summary \& Conclusions}

NGC~7673 is a fascinatingly disturbed galaxy, with a total SFR 
($\approx 10-20$~M$_{\odot}$~yr$^{-1}$) and a SF intensity of 
for Clump~B ($\sim 1$~M$_{\odot}$yr$^{-1}$kpc$^{-2}$) that put it in
the range
populated by Lyman break galaxies (\cite{Petetal01}). Due to its combination
of proximity for a galaxy with CNELG properties and favorable face-on
orientation, it affords an excellent opportunity to explore the structure
of a major starburst. While further study is needed to fully disentangle
the effects of age and reddening on star cluster colors, our initial
investigation reveals several interesting trends.

1. Working from WFPC2 images, we have identified 50 star cluster
candidates where we measured F255W, F555W and F814W magnitudes, and 268
with F555W and F814W photometry. While we find a broad range of star
cluster colors, the bluer and brighter clusters are strongly concentrated
into the main starburst `clumps'.

2. In our three color sample we mainly measure clusters with ages of
$<$20~Myr. 
All of the main optical clumps, A, B, C, D, and F contain star
clusters with ages of $5$~Myr or younger. However, the nuclear region,
clump A, may contain older clusters than are found in clumps B and C. Even
so we do not find an obvious age ordering of the clumps. Further
observations with more filters and deeper data are needed to allow better
separation of clusters which are red due to age versus young clusters
reddened by interstellar dust.

3. The four brightest clumps are composed of $> 50$ star clusters detected
from the UV through the optical, providing $\sim16-33\%$ of the luminosity
from the clump as a whole. These results are consistent with other
estimates for the fraction of light contributed by compact star clusters
in starbursts (e.g., \cite{Metal95})

Clump~F is distinguished by not being an association of star clusters.
It is instead dominated by a single object, for which we derive an age of
$4-5$~Myr and a mass of $3 \times 10^{5}$~M$_{\odot}$. If the age is 4~Myr or
slightly less, it should show the presence of Wolf-Rayet stars in its
spectrum. Perhaps Clump~F resembles a more luminous version of compact
young stellar complexes with a dominant super star cluster, such as the
Hodge complex in NGC~6946 (\cite{Laretal02}) or the extended knot S that is
composed of luminous young stars in the Antennae (\cite{Wetal99}).

4. The star formation process in NGC~7673 differs from that seen in normal
spirals in that the disk has broken up into relatively well-defined clumps
that contain most of the current star-forming activity. The clumps may be
understood as the result of large spatial scale instabilities in perturbed
gaseous disks, as suggested by Elmegreen et al. (1993). 
 However, it is not yet clear if the current clumps can survive
long enough for dynamical friction to act to bring clumps
into the center of the galaxy where they can form a bulge, as suggested
by Noguchi (1999).

The unusual structure of NGC~7673 and other nearby starburst galaxies
indicate that the impact of highly coeval populations of star clusters
should be taken into account when considering how best to model galaxies
with intense star formation. These issues extend from implications for
the integrated rest-frame UV spectra to their impact on the ISM, and even
to the subsequent evolution of the galactic disks (\cite{K02b}). Furthermore,
we must take into account that the formation of dense star clusters is 
a signature of efficient star formation, where $\ge 20 \%$ of the gas 
is converted into stars (e.g. \cite{H80}, \cite{L99}). Formation 
of numerous dense clusters can
allow a galaxy to more efficiently convert gas into stars, thereby 
sustaining high SFRs in starburst systems.

\acknowledgements 

This paper was improved by careful comments from the referee, Danielle
Alloin. N. H. acknowledges the ESO Studentship Programme
and the Wisconsin Space Grant Consortium Graduate Fellowship Program.
J. S. G. and N. H. thank the University of Wisconsin Graduate School for
partial support. N. H. would also like to thank Ritter Sport for many 
bright, chocolatey spots in many gray Garching days, and S.B. for additional 
encouragement and support.
This project uses archival HST WFPC2 data obtained by the WFPC2
Investigation Definition Team for studies of luminous blue galaxies, and
we thank the IDT for its early support of this effort.

\end{document}